\documentclass{article}
\usepackage{spconf,amsmath,graphicx}

\usepackage{amssymb}
\usepackage{booktabs}
\usepackage{multirow}
\usepackage{hyperref}

\title{Improving PSEUDO-LABEL training For End-to-end Speech Recognition Using Gradient Mask}
\name{Shaoshi Ling, Chen Shen, Meng Cai, Zejun Ma}
\address{Bytedance AI Lab\\
lingshaoshi@bytedance.com}
\begin{document}
%
\maketitle
\begin{abstract}
In the recent trend of semi-supervised speech recognition, both self-supervised representation learning and pseudo-labeling have shown promising results. In this paper, we propose a novel approach to combine their ideas for end-to-end speech recognition model. Without any extra loss function, we utilize the \textit{Gradient Mask} to optimize the model when training on pseudo-label. This method forces the speech recognition model to predict from the masked input to learn strong acoustic representation and make training robust to label noise. In our semi-supervised experiments, the method can improve the model's performance when training on pseudo-label and our method achieved competitive results comparing with other semi-supervised approaches on the Librispeech 100 hours experiments. 


\end{abstract}
\begin{keywords}
speech recognition, semi-supervised learning, pseudo-labeling, end-to-end model
\end{keywords}
\section{Introduction}
\label{sec:intro}

Pseudo-labeling \cite{thomas2013deep, huang2016semi, manohar2018semi, karita2018semi, parthasarathi2019lessons, kahn2020self, xu2020iterative, park2020improved, likhomanenko2020slimipl} is one of the most popular semi-supervised learning approaches and recently demonstrated its efficacy in automatic speech recognition. In this approach, a smaller labeled set is used to train an initial seed model, which is applied to a larger amount of unlabeled data to generate hypotheses. The unlabeled data with the most reliable hypotheses are added to the training data for re-training. This process can be repeated iteratively to improve the quality of pseudo labels \cite{xu2020iterative}. However, pseudo-label training is sensitive to the quality of the hypotheses. Errors or noise in labels can cause training unstable and resulted in sub-optimal states, especially for end-to-end speech recognition models \cite{kahn2020self}. Thus, pseudo-label training usually requires a careful calibration by the confidence measures \cite{kahn2020self, park2020improved}. But confidence-based data filtering will not always work perfectly since most pseudo-label sequences would contain errors. 



Starting from BERT \cite{devlin2019bert}, masked prediction has becoming a new principle to solve problems in self-supervised settings in NLP. The core idea of masked prediction is to force the model to learn good high-level representations of unmasked inputs to infer the targets of masked ones correctly. In speech, the approaches sharing the same spirit have been proposed: masked prediction of audio acoustic features \cite{liu2020mockingjay, liu2020tera}, masked prediction of quantized acoustic features \cite{ling2020decoar} and masked prediction of unsupervised clusters \cite{hsu2021hubert}. Experiments in \cite{hsu2021hubert} also showed that computing loss only from the masked regions achieves better performance than all regions. 




We draw inspiration from masked prediction and integrate its idea into pseudo-label training. We propose the \textit{Gradient Mask} to improve pseudo-label training in end-to-end speech recognition. In our approach, we first train a seed model to generate pseudo labels and then use the Gradient Mask to train a student model on the pseudo labels. The model only allows gradients corresponding to masked input back-propagate through the model encoder by masking the gradients corresponding to unmask input. The model is trained by jointly minimizing the loss on labeled and pseudo-label data while the Gradient Mask is turned off on labeled data. 

Our training method can force the model to learn strong acoustic representation in order to infer from masked input. Moreover, it can also improve pseudo-label training by making the model less affected by label noise. The intuition is that only gradients of the masked part are used when updating the model's parameters, so it can avoid the sudden dramatic change in gradients caused by errors and also alleviate the overfit to corrupted labels. Our approach is simple and efficient since it doesn't require any extra parameters, extra loss or data filtering steps. We run our experiments using the Transducer model \cite{graves2012sequence}. The experiment showed that our method is robust to label noise and can achieve competitive results comparing with other self/semi-supervised approaches in the Librispeech 100 hours experiments.

\section{Related work}
\label{sec:format}

\subsection{Combating noisy labels}
DNNs are known to be susceptible to noisy labels \cite{frenay2013classification, song2020learning} and the errors in labels could be extremely harmful to models. Beyond conventional data filtering/cleaning techniques, deep learning techniques have recently gained vast interest. There are several works to investigate supervised learning under noisy labels \cite{song2020learning} in computer vision. However, these models cannot be directly applied to ASR and fewer studies have proposed to combat noise labels for ASR. In \cite{hasegawa2016asr}, the phonetic sequence was inferred from several noisy transcriptions made by non-native transcribers using a misperception model, and then used to train a conventional hybrid ASR model. \cite{dufraux2019lead2gold} propose a novel loss function that jointly learns the ASR model and a transcription graph that can search for better transcriptions of the training data.

\subsection{Joint training with self-supervised and ASR tasks}
The idea of self-supervised learning \cite{liu2020mockingjay, liu2020tera, ling2020decoar, hsu2021hubert, schneider2019wav2vec, chung2019unsupervised, ling2020deep, baevski2020wav2vec} is to learn speech representations that are useful for ASR. By first pre-training on a large amount of unlabeled data using a proxy task, the model can be fine-tuned on labeled data and achieved impressive results. This process is a two-stage process as it requires running separate pre-training and fine-tuning. Joint training with speech recognition and self-supervised representation \cite{ling2020bertphone, talnikar2021joint, wang2021unispeech, wang2021unispeech2} is the line of work to simplify this process and is the closest to our method. Those methods typically have two training objectives: one is for the ASR task on the labeled data while the other is to train self-supervised representation (e.g. masked feature prediction \cite{ling2020bertphone}), on the unlabeled data. Our method is much simpler and uses only one loss on both label and unlabeled data (pseudo-label data). 



\section{Method}
In speech recognition, an E2E model predicts the conditional distribution $P(Y|X)$ of token sequences $Y = \left[ y_1, . . . , y_U \right]$ given a speech-feature sequence $X = \left[x_1, . . . , x_T \right]$ as the input, where $y \in V$ and $x_t$ is acoustic feature vectors at time t. V is the set of all possible output tokens. We will explain and show our method in transducer model \cite{graves2012sequence}, but it can be perfectly adapted to other end-to-end ASR models (e.g. CTC \cite{graves2016ctc}, seq2seq \cite{chan2016listen}) as well.

\subsection{Transducer model} \label{transducer}
Transducer model \cite{graves2012sequence} consists of encoder, prediction network and joint network. The encoder $f_{enc}$ encode the inputs X to higher-level representation $h_{enc} = \left[ h_{1}^{enc}, ..., h_{T}^{enc} \right]$. 

The prediction network takes embedding vectors of
previous non-blank labels $ \left[ y_{1}, ..., y_{u-1}  \right]$ as input to produce its output $h^{pred}_u$ at step $u$. Then the logits over vocabulary at frame $t$ and step $u$ can be computed by the joint network:

\begin{equation}\label{jointer}
  \begin{aligned}
h^{joint}_{t,u} = f_{joint}(h^{enc}_t, h^{pred}_u)
\end{aligned}
\end{equation}

The probability distribution over vocabulary at
frame $t$ and step $u$ is calculated using a soft-max layer. With
forward-backward algorithm, the sum probability $P(Y|X)$
of all alignment paths is adopted as the objective function.

\subsection{Gradient mask} \label{gm}
For sequence $X = \left[x_1, . . . , x_T \right]$ which has pseudo labels $ {Y}' = \left[ {y}'_{1}, . . . , {y}'_{u-1} \right]$. The objective is to enable model to predict the labels from the masked features. In another word, $f_{enc}$ is trained to be a strong acoustic representation model which can benefit the ASR tasks.

Before feeding features to the encoder, we randomly generated a sequence $mask = \left[m_1, . . . , m_T \right]$ representing the mask positions for the input sequence $X$. Specifically, ${m_t}$ is 1 if features are masked at time $t$, otherwise ${m_t}$ is 0. The features, for example $x_t$, are masked by replacing it with a learnt mask embedding $emb$. Then the encoder $f_{enc}$ encode this mask sequence as :
\begin{equation}
  \begin{aligned}
h^{enc} = f_{enc}((\sim mask) * X + mask * emb)
\end{aligned}
\end{equation}

Our mask strategy is the same as \cite{baevski2020wav2vec}, where we randomly sample without replacement a certain proportion $p$ of all time steps to be starting indices and then mask the subsequent $m$ consecutive time steps from every sampled index with overlap spans.

When the gradient is back-propagated to the encoder, we masked the gradients corresponding to the non-masked inputs using $mask$ sequence:
\begin{equation}
  \begin{aligned}
grad_{h^{enc}} = (\sim mask) * grad_{h^{enc}}
\end{aligned}
\end{equation}

And the prediction network takes the pseudo labels sequence as the input. The joint network then produces output as in \eqref{jointer}. But when we do back-propagation, we also block the gradient flow into the predictor network. This process can be expressed in the following functions:

\begin{equation}
  \begin{aligned}
h^{joint}_{t,u} = f_{joint}(h^{enc}_t, sg( h^{pred}_u))
\end{aligned}
\end{equation}

where $sg(x) \equiv  x$, $\frac{d}{dx}sg(x) \equiv  0$ is the stop gradient operator. The objectice function is still the same transducer loss where we trying to minimize $P({Y}'|X)$ of all alignment paths.


\begin{table*}[th]
    \centering
   \begin{tabular}{c|c|c|c|c|cc|cc}
        \toprule
        \multirow{2}{*}{\textbf{Method}} &
        \multirow{2}{*}{\textbf{Model Size}} &
        \multirow{2}{*}{\textbf{Criterion}} &
        \multirow{2}{*}{\textbf{LM}} &
        \multirow{2}{*}{\textbf{G/TPU-days}} &
        \multicolumn{2}{c|}{\textbf{dev}} & \multicolumn{2}{c}{\textbf{test}} \\
         & &&& &clean & other & clean & other \\
        \midrule
        NST \cite{park2020improved} & 360M & S2S & lstm & 1600 &3.9 &8.8& 4.2& 8.6 \\
        w2v2-base\cite{baevski2020wav2vec} & 95M & CTC & None &  102.4 & 6.1 &13.5& 6.1& 13.3 \\
        w2v2-large\cite{baevski2020wav2vec} & 317M & CTC & None &  294.4 &4.6 &9.3 & 4.7 &9.0 \\
        IPL \cite{xu2020iterative} & 322M & CTC & None& 192  & 5.5 &9.3 &6.0 &10.3 \\ 
        slimIPL \cite{likhomanenko2020slimipl} & 322M & CTC & None&   83.2 &3.7 &7.3 & 3.8  & 7.5 \\ 
        \midrule
        NST-iter1 &  118M & transducer & None & 14 & 5.3 & 12.7& 5.4 & 12.9 \\
        GM-iter1 &  118M & transducer & None & 14 & 4.8& 11.1 & 4.9 & 11.2 \\
        GM-iter5 &  118M & transducer & None& 54 &4.1 & 8.8 &4.3 & 8.8 \\
        \bottomrule
    \end{tabular}
    \caption{Semi-supervised LibriSpeech results using 100 hours as labeled data and 860 hours as unlabled data. Our experiments are in the lower part of the table.}
    \label{table:LibriSpeech}
\end{table*}

\subsection{Training procedure}
The whole training process is similar to the standard pseudo-labeling approach. Let $L =\{x_i, y_i\}$ be a labeled dataset and $ U =\{x_j\}$ be a large unlabeled dataset. We first train a seed acoustic model M on the labeled dataset $L$. We use this seed acoustic model M to generate pseudo-labeled on dataset $U = \{x_j, {y}'_j\}$ and we then combine it with all the label data in L to form new dataset ${U}'$ = $U \cup L$. 

The next step is to train a student model using both the datasets $L$ and ${U}'$. The model is trained by alternately minimizing the losses on $L$ and ${U}'$. When updating the model parameters using a minibatch from the pseudo-labels dataset ${U}'$, we apply the gradient mask method as described in \ref{gm} on the model. While on a minibatch from the labeled dataset, we do parameters update in the standard way for transducer in \ref{transducer}. This process is repeated until convergence of the word error rate on the validation dataset. Since the loss function is the same for both datasets, we only use one momentum optimizer and the same learning rates for simplicity. The ratio of minibatch from $L$ to minibatch from ${U}'$ is a hyper-paramtete to be tuned.


\begin{table}[thb]
    \centering
    \ninept
    \begin{tabular}{c|c|cc|cc}
        \toprule
        \multirow{2}{*}{\textbf{Method}} &
        \multirow{2}{*}{\textbf{LM}} &
        \multicolumn{2}{c|}{\textbf{dev}} & \multicolumn{2}{c}{\textbf{test}} \\
        & &clean & other & clean & other \\
        \midrule
        Hybrid \cite{luscher2019rwth} & 4-gram & 5.0 & 19.5 & 5.8 &18.6 \\
        LAS \cite{park2020improved}  & lstm & 5.3 &16.5& 5.5 &16.9\\
        CTC \cite{likhomanenko2020slimipl}  & None & 6.2&16.8& 6.2 &16.8 \\
        \midrule
        Transducer & None &  6.3& 16.8 & 6.4 & 16.7\\
        \bottomrule
    \end{tabular}
    \caption{WER on the Librispeech 100 hours for supervised system}
    \label{table:baseline}
\end{table}

\section{Experiments}
\subsection{Data}
We conducted our experiments on the LibriSpeech \cite{panayotov2015librispeech} datasets. The labeled dataset is a 100 hours subset (train-clean-100) of Librispeech, and the remaining 860 hours (train-clean-360, train-other-500) is the unlabeled dataset. During training, samples in the dataset that are longer than 20 seconds are filtered out. The performance of the trained model is validated on the dev-clean and dev-other datasets of Librispeech and tested on the test-clean/other dataset. We did not use any extra text or LM information for any of our experiments.

We use around 5k subword \cite{sennrich2015neural} units as our prediction targets. We extracted 80-channel filterbank features computed from a 25ms window with a stride of 10ms. When training on labeled data, we use speed perturbation and SpecAugment \cite{park2019specaugment, gulati2020conformer} with mask parameter (F = 27), and ten time masks with maximum time-mask ratio (pS = 0.05), where the maximum-size of the time mask is set to pS times the length of the utterance.

\subsection{Setup}

The filterbank features are first passed into 2 blocks of 2d-conv layers, time reduction layers are added after each block to down-sample the frame rate to 4 before passing into the encoder. The encoder model consists of 17 layers of conformer block, where we set the model dimension to 512, the inner dimension in feed forward layer to 2048, with 8 attention heads, 32 kernal size in convolution block, with the same setting as Conformer-L \cite{gulati2020conformer}. We use LSTM as our predictor and the LSTM predictor contains 1 layer with 640 units and a projection layer with 640 units. The Transducer's joint network is a simple feed-forward layer. The total number of parameters is about 130M. Our model is implemented in Pytorch and we optimized our model using Adam. We use this same model in all of our experiments.

For the 100 hours seed model, we first train the GMM-based model in Kaldi \cite{povey2011kaldi} to obtain the alignment results on the 100 hours subset, and we use the frame-wise phoneme label to pre-train the encoder. Then we use the pre-trained encoder to initialize our transducer model \cite{hu2020exploring}. For training the transducer model, we use learning rate warm-up for the first 10k updates to a peak of 1e-4, and hold for 60k steps, then linearly decayed it. We grouped the input sequences by length with a batch size of 10k frames per GPU, and trained the models on 4 GPUs for 160k steps in total. 

For training the student model, the mask $p$ is set to 0.065 and $m$ is set to 3 (equal to 12 frames or 0.12 second). This masking schema is similar to \cite{baevski2020wav2vec}, and it resulted in around half of frames being masked. We set the ratio of minibatch from labeled data to pseudo-label data to 1:9. This ratio is the same as the ratio of amount of data and it produces an ASR model with the best performance. We used learning rate warm-up for the first 10k updates to a peak of 2e-4, and hold for 80k, then linearly decayed it. We grouped the input sequences by length with a batch size of 10k frames per GPU, and trained the models on 8 GPUs for 180k steps.

\subsection{Results}
\label{sec:expt}
\subsubsection{Supervised baseline}
Table \ref{table:baseline} shows the results of our seed model and the comparison with the Librispeech 100 hours supervised model from other papers. We use this seed model to generate the first version of the pseudo-label. The resulted 860 hours pseudo-label have WER around 9.

\subsubsection{Semi-supervised experiments}
Table \ref{table:LibriSpeech} shows the results from semi-supervised experiments. \textit{NST-iter1} is the results of the experiment where we simply mixed the pseudo-label data and labeled data to form the new training dataset, and we train the student model using this dataset. This process is a simplified version of noise student training since we did not do any filtering, LM fusion, or data selection \cite{kahn2020self, park2020improved}. 

\textit{GM-Iter1} and \textit{GM-Iter5} is the model using gradient mask method. For \textit{GM-Iter1} in the table are the results from the student model directly trained from the pseudo labels generated by the seed model. Our proposed approach significantly outperforms the 100 hours supervised baselines in table \ref{table:baseline} and also the noisy student training baseline. For \textit{GM-Iter5}, we iterate the pseudo labeling process 5 times. In particular, the model of \textit{GM-Iter5} achieved highly competitive performance, with a WER of 4.1/8.8 for dev-clean/dev-other and 4.3/8.9 for test-clean/test-other. It is worth noting that our method is highly efficient. We use much fewer computing resources or a much smaller model size compared with other approaches in the table \ref{table:LibriSpeech}.

\subsection{Ablation study and analysis}
\subsubsection{Pseudo-labeling iterations}
To study the performance from different pseudo-labeling iterations. The table \ref{table:iterations} shows the WER on test-clean/other of each training iterations using the gradient mask method. The results are in table \ref{table:iterations}. We stopped this process after the 5th iteration since the improvement is already minimum at the iter5. 

\begin{table}[thb]
    \centering
    \ninept
    \begin{tabular}{ c | cc  }
        \toprule
        {\textbf{interations}} & {\textbf{test clean} } & {\textbf{test other} } \\
        \midrule
         100h seed & 6.2 &16.8 \\
         iter1 & 4.9 & 11.2 \\
         iter2 & 4.6 & 9.7\\
         iter3 & 4.4 & 9.2\\
         iter4 & 4.3 & 8.9\\
         iter5 & 4.3 & 8.8\\
        \bottomrule
    \end{tabular}
    \caption{Ablation on each pseudo-labeling iterations}
    \label{table:iterations}
\end{table}

\subsubsection{Gradient mask on labels of different qualities}
We conduct an ablation study to investigate the effect of the gradient mask on labels of different qualities. We run the experiments with and without the gradient mask method on those labels. The training is the same as the standard transducer training when we do not use the gradient mask. The training data includes 860 hours pseudo-label data and the 100h labeled data. Pseudo(WER-9) is the pseudo-label generated by the 100h seed model which has around WER 9. Pseudo(WER-15) is generated by the same supervised system but from an early epoch that has WER around 15. Pseudo(WER-5) is generated by the student model from the 3rd iteration. And Pseudo(WER-2) is generated by an intermediate model trained on 960 hours labeled data.

When pseudo-label contains a lot of errors (WER 15), simply adding pseudo label will cause the model performance to degrade comparing with 100h baseline in table \ref{table:baseline}. Even when we have high-quality pseudo-label (WER 5), the noise in labels still hurts the model performance. On the other hand, the gradient mask method can be robust to bad quality labels and work well consistently on the label of different quality. We found that the worse pseudo-label's quality, the better performance we can obtain using the gradient mask method comparing with the standard training. The standard training will perform comparably to the gradient mask method when pseudo-label has WER around 2, and it would perform better when we use the ground truth reference labels.


\begin{table}[thb]
    \centering
    \ninept
    \begin{tabular}{ c | cc  }
        \toprule
        {\textbf{data}} & \multicolumn{2}{c}{\textbf{dev-other}} \\
        & gm & w/o gm\\
        \midrule
        Reference & 7.5 & 6.7 \\
        Pseudo(WER-2) & 7.8 & 7.6 \\
        Pseudo(WER-5) & 8.8 & 9.4 \\
        Pseudo(WER-9) &  11.1 & 12.7\\
        Pseudo(WER-15) & 14.2 &  18.9\\
        \bottomrule
    \end{tabular}
    \caption{With and without gradient mask on different pseudo-label}
    \label{table:vq_layer}
\end{table}

\section{conclusion}
In this paper, we present the Gradient Mask method, a simple and efficient method to improved pseudo-label training for end-to-end speech recognition. Our method can force the model to learn acoustic representation and also be robust to errors in labels. This method can be used to combat label noise in pseudo-label training. In semi-supervised experiments, our method achieved much better performance than the conventional pseudo label training approach and performed comparably to the SOTA approach while being much computation-efficient. Future work includes exploring the extension to other end-to-end ASR systems like LAS and other sequence to sequence tasks like machine translation. 

\footnotesize
\bibliographystyle{IEEEbib}
\bibliography{refs}

\end{document}